\documentclass[12pt,fleqn]{article}

\usepackage{epsfig}
\usepackage{graphics}
\usepackage{latexsym}
\usepackage{amssymb}
\usepackage{amsmath}
\newcommand{\creation}[1]{\lambda_{#1}}
\newcommand{\anihilation}[1]{\frac{\partial}{\partial \lambda_{#1}}}
\newcommand{\derivation}[1]{\frac{\partial}{\partial q_{#1}}} 
\newcommand{\ben}{\begin{displaymath}}
\newcommand{\een}{\end{displaymath}}
\newtheorem{lemma}{Lemma}
\newtheorem{theorem}[lemma]{Theorem}

\newtheorem{corollary}[lemma]{Corollary}

\title{Vanishing index for supersymmetric 2-matrix model with odd dimensional
gauge group}
\author{G.M. Graf${}^{(a)}$, D. Hasler${}^{(a)}$,  J.
Hoppe${}^{(b)}$ \\ 
\vspace*{-0.05truein} \\
\normalsize\it ${}^{(a)}$ Theoretische Physik,
ETH-H\"onggerberg, CH--8093 Z\"urich\\
\normalsize\it   ${}^{(b)}$ Royal Institute of Technology, Department of
Mathematics, S-10044 Stockholm }

\begin{document}

\maketitle
\vspace{0.4cm}
\begin{abstract}
We define an operator which for odd-dimensional compact gauge group furnishes
 unitary equivalence of the bosonic and fermionic sector in the supersymmetric
quantum-mechanical matrix model obtained by dimensional reduction from
3-dimensional 
supersymmetric Yang-Mills theory. 
\end{abstract}

Let $G$ be a compact semisimple Lie group. We consider a class of
supersymmetric  $G$-invariant quantum mechanical models involving $d=2$ copies
of the Lie algebra of $G$. If the dimension $g$ = dim$G$ of $G$ is
odd, we will show that the 
Hodge $*$ operator, in a simple, direct way, gives rise to a unitary
equivalence of the Hamiltonian restricted to the bosonic and fermionic
sector. This
trivially implies the vanishing of the Witten-index (see [1-7] for work on
the subject). 

Let us first describe the Hilbert space. The
configuration 
space of the bosonic degrees of freedom is $X = \mathbb{R}^{2 g}$, with
coordinates 
\ben
(q_{sA})_{
s = 1,2, \, 
A = 1,\ldots,g} .
\een
To describe the fermionic degrees of freedom one considers the Clifford
algebra with hermitian generators $(\Theta_{\alpha A})_{\alpha=1,2,\,
A=1,\ldots,g}$ and commutation relations
\ben 
\left\{ \Theta_{\alpha A}, \Theta_{\beta B} \right\} = \delta_{\alpha \beta} \delta_{A
B}  . 
\een 
We realize the Clifford algebra  on the fermionic Fock
space $\Lambda$
over the Hilbert space $\mathbb{C}^{g}$. We decompose
\ben
\Lambda = \bigoplus_{p=0} \Lambda_p \ , \ \ \Lambda_+ = \bigoplus_{p=0}
\Lambda_{2p} \ , \ \ \Lambda_- = \bigoplus_{p=0} \Lambda_{2p + 1}  ,
\een
where $\Lambda_p = {\bigwedge}^p(\mathbb{C}^g)$ denotes the $p$-particle 
space. $\Lambda$ carries a natural scalar product
$\langle \cdot , \cdot \rangle$. Let
\ben
(e_{A})_{A=1,\ldots,g}
\een
be the standard basis in $\mathbb{C}^g$, and $\creation{ A}$,
$\anihilation{A}$ the associated creation and annihilation
operators. The
Clifford generators can now be expressed as  
\ben 
\Theta_{1 A}=\frac{1}{\sqrt{2}}\left(\lambda_{ A}+    
\frac{\partial}{\partial \lambda_{ A}}\right), \   
\Theta_{2   A}=\frac{i}{\sqrt{2}}\left(\lambda_{ A}-  
\frac{\partial}{\partial \lambda_{ A}} \right). 
\een
Let $(-1)^{F} = {\left( \frac{i}{2} \right)}^g \prod_{A} \Theta_{1 A}\Theta_{2 A}$,
which equals $\pm 1$ on  
$\Lambda_{\pm}$. We 
define the Hilbert spaces
\ben
\mathcal{H} = \mathrm{L}^2(X) \otimes \Lambda = \mathcal{H}_+ \oplus
\mathcal{H}_-  ,
\een
with $\mathcal{H}_{\pm} = \mathrm{L}^2(X) \otimes \Lambda_{\pm}$, called the bosonic and
fermionic sector, respectively. The Hilbert space $\mathcal{H}$ carries a unitary
representation of $G$. Let $\mathcal{H}^{(0)}$ denote the physical
Hilbert space consisting of $G$-invariant states in $\mathcal{H}$. $\mathcal{H}^{(0)}_{+},
\mathcal{H}^{(0)}_{-}$ are given analogous. We define the Hodge $\ast$
operator with
respect to the volume form $\omega = e_1 \wedge \ldots \wedge e_g$ in $\Lambda$, i.e.
\ben
\ast : \Lambda \rightarrow \Lambda :
\een
for $\beta \in \Lambda_p$, let $\ast \beta \in \Lambda_{m - p}$ be the unique
element such that
\ben
\langle \ast \beta , \alpha \rangle = \langle \omega , \overline{\beta} \wedge
\alpha \rangle  , \, \mathrm{for \ all} \ \alpha \in \Lambda_{m-p} .
\een
By $\overline{\beta}$ we mean complex conjugation of $\beta$.
We list some properties of the Hodge
$\ast$ in the following lemma, which can easily be verified.
\begin{lemma} \
\begin{description}
\item[(a)] ${\ast}^{\dagger} = {\ast}^{-1}$
\item[(b)] On $\Lambda_p$ we have, ${\ast}^2 = (-1)^{(m-p)p}$.
\item[(c)] If $m$ is odd, then $\ast : \Lambda_+ \rightarrow \Lambda_-$ and
$\ast : \Lambda_- \rightarrow \Lambda_+$, i.e. $\{ \ast , {(-1)}^F \} = 0$.
\item[(d)] On $\Lambda_p$ we have, $\lambda_{A} \ast = {(-1)}^{p-1} \ast
\anihilation{A}$ and $\anihilation{A} \ast = {(-1)}^p \ast \lambda_{A}$.
\end{description}
\end{lemma}

The model is defined
by the following operators. Let $f_{ABC}$ be the real and totally antisymmetric
structure constants, with respect to an orthonormal basis of the Lie algebra of
$G$. The generators of the unitary representation on $\mathcal{H}$ are
\ben
L_A = - i f_{ABC} \left( q_{sB} \derivation{sC} + \lambda_B \anihilation{C}
\right) , \, A=1,\ldots,g
\een
The supercharges are given by
\begin{eqnarray*}
Q & = & \left( \derivation{1A} - i \derivation{2A} \right) \anihilation{A} + i f_{ABC}
q_{1A} q_{2B} \lambda_{C}  , \\
Q^{\dagger} & = & - \left( \derivation{1A} + i \derivation{2A} \right)
\lambda_A - i f_{ABC} q_{1A} q_{2B} \anihilation{C} \ 
\end{eqnarray*}
and the Hamiltonian by
\ben
H = - \Delta + \frac{1}{2} {\left( f_{ABC} q_{mB}q_{nC} \right)}^2 + \left(
q_{1A} + i q_{2A} \right) f_{ABC} \lambda_{B} \lambda_{C}  -
\left( q_{1A} - i q_{2A} \right) f_{ABC} \anihilation{B} \anihilation{C} \ .
\een 
The physical Hilbert space can be written as
\ben
\mathcal{H}^{(0)} = \left\{ \psi \in \mathcal{H} \mid  L_A \psi = 0, \,
A=1,\ldots,g \right\}  . 
\een
We recall the super-algebra
\begin{eqnarray*}
&&\left\{ Q , Q^{\dagger} \right\} {\mid}_{\mathcal H^{(0)}}  =   H
{\mid}_{\mathcal H^{(0)}} , \ \ 
\left\{ Q , Q \right\} {\mid}_{\mathcal H^{(0)}}  =  0 , \ \ 
\left\{ Q^{\dagger}, Q^{\dagger} \right\} {\mid}_{\mathcal H^{(0)}}  = 0 ,  
\\
&&\left\{ {(-1)}^F , Q \right\}  =   0  , \ \   \left\{ {(-1)}^F , Q^{\dagger}
\right\} = 0   
 .
\end{eqnarray*}
We define the following operators $S$ and $P$ by
\begin{eqnarray*}
&S& : \mathrm{L}^2(X)  \rightarrow  \mathrm{L}^2(X) \\
&  &  \ \ \   \ \ \ \psi  \mapsto  S\psi(q_{1A},q_{2A}) = \psi(q_{1A},-q_{2A})  , \\
&P&=  S \ast = \ast S  .
\end{eqnarray*}
$P$ has the following properties.
\begin{theorem} \ 
\begin{description} 
\item[(a)] $P$ is unitary. 
\item[(b)] If  dim $G$ is odd, then we have
$P^2=1$ and 
$P:\mathcal{H}_+ \to \mathcal{H}_-, \ P:\mathcal{H}_- \to \mathcal{H}_+$,
i.e. $\{ P , {(-1)}^{F} \} = 0$. 
\item[(c)] $[P,H] = 0 \ , \ \ [P,L_A] = 0$.
\item[(d)] On $\mathcal{H}_-$ we have $QP = PQ^{\dagger}$, $Q^{\dagger} P
= - P Q$ and on $\mathcal{H}_+$, $QP = - PQ^{\dagger}$, $Q^{\dagger}P = P Q$.
\end{description}
\end{theorem}
\noindent  
{\it Proof of Theorem.\  }  
     
(a) follows by $S^{\dagger}=S^{-1}$ and Lemma 1(a). 

(b) follows from $S^{-1}=S$ and Lemma 1(b),(c).

(c) Applying Lemma 1(d), we find
\ben
\ast \lambda_B \lambda_C  = - \anihilation{B} \anihilation{C} \ast , \ \ 
\ast  \anihilation{B} \anihilation{C} = -   \lambda_B \lambda_C \ast , \ \
\ast   \anihilation{B} \lambda_{C} = \lambda_{B}  \anihilation{C} \ast  ,
\een
which shows that $[P,H]=0$ and $[P,L_A]=0$.

(d). Applying Lemma 1(d), we find
\begin{eqnarray*}
\mathrm{on} \ \Lambda_+ & : & \lambda_A \ast =  - \ast \anihilation{A}, \ 
\anihilation{A} \ast =  \ast \lambda_A, \\
\mathrm{on} \ \Lambda_- & : & \lambda_A \ast =  \ast \anihilation{A}  , \ 
\anihilation{A} \ast = - \ast \lambda_A  \ ,
\end{eqnarray*}
from which (c) follows. \ \ $\Box $  \\
This theorem directly implies the following corollary.
\begin{corollary} 
Let dim $G$ be odd.  
\begin{description}  
\item[(a)] $P: \mathcal{H}_+^{(0)} \rightarrow \mathcal{H}_-^{(0)}$,       
$P: \mathcal{H}_-^{(0)} \rightarrow \mathcal{H}_+^{(0)}$. 
\item[(b)] $P( H     
{\mid}_{\mathcal{H}_+^{(0)}}) P = H {\mid}_{\mathcal{H}_-^{(0)}}$,      
i.e.  $H {\mid}_{\mathcal{H}_+^{(0)}}$ and $H {\mid}_{\mathcal{H}_-^{(0)}}$
are      
unitarily        
aequivalent.          
\end{description}         
\end{corollary}      
The corollary implies that the Witten index vanishes. In fact, both the
principle and deficit contribution to the Witten index (cp. [2-7])
vanish, since $P$ 
commutes with the characteristic function of the ball $B_R(0) = \{ x : \| x \|
\leq R 
\}$ for any $R$.

\medskip  
\noindent {\bf Acknowledgments.\/} We thank J. Fr\"ohlich for valuable discussions.


\begin{thebibliography}{99}
 



\bibitem{fh} J. Fr\"ohlich, J. Hoppe, On zero-mass ground states in 
super-membrane matrix models. Comm. Math. Phys. {\bf 191}, (1998) 613-626  
 
   
\bibitem{y} P. Yi, Witten Index and threshold bound states of D-Branes, Nucl
Phys. B {\bf 505 }, (1997) 307 

\bibitem{ss} S. Sethi, M. Stern, Comm. Math. Phys. {\bf 194}, (1998) 675


\bibitem{kns} W. Krauth, H. Nicolai, M. Staudacher, Phys. Lett. {\bf B431},
(1998) 31 

\bibitem{gg} M.B. Green, M. Gutperle, D-particle bound states and the
D-instanton measure, JHEP 9801 (1998) 005

\bibitem{mns}  G. Moore, N. Nekrasov, S. Shatashvili, D-particle bound states
and generalized instantons, Comm. Math. Phys. {\bf 209}, (2000) 77-95

\bibitem{ks} V.G. Kac, A.V. Smilga, Normalized Vacuum States in N = 4 
Supersymmetric Yang--Mills Quantum  
Mechanics with Any Gauge Group, Nucl.Phys. B {\bf 571}, (2000)  515-554 
 




\end{thebibliography}
\end{document}